\documentclass[12pt,preprint]{aastex}

\shorttitle{Carbon Planets}
\shortauthors{Kuchner}

\begin{document}

\title{Extrasolar Carbon Planets}

\author{Marc J. Kuchner\altaffilmark{1}}
\affil{Princeton University Department of Astrophysical
Sciences \\
Peyton Hall, Princeton, NJ 08544}

\author{S. Seager}
\affil{Carnegie Institution of Washington,
5241 Broad Branch Rd. NW, Washington DC 20015}
\email{seager@dtm.ciw.edu}

\altaffiltext{1}{Hubble Fellow}

\begin{abstract}

We suggest that some extrasolar planets $\lesssim
60$~M${}_{\bigoplus}$ will form substantially from silicon carbide
and other carbon compounds.   Pulsar planets and low-mass white dwarf
planets are especially good candidate members of this new class of
planets, but these objects could also conceivably form around stars
like the Sun.  This planet-formation pathway requires only a factor of
two local enhancement of the protoplanetary disk's C/O ratio above
solar, a condition that pileups of carbonaceous grains may create in
ordinary protoplanetary disks.  Hot, Neptune-mass carbon planets
should show a significant paucity of water vapor in their spectra
compared to hot planets with solar abundances.  Cooler, less massive
carbon planets may show hydrocarbon-rich spectra and tar-covered
surfaces.  The high sublimation temperatures of diamond, SiC, and
other carbon compounds could protect these planets from carbon
depletion at high temperatures.

\end{abstract}


\section{INTRODUCTION}

The recent discoveries of Neptune-mass extrasolar planets by the
radial velocity method \citep{sant04, mcar04, butl04} and the rapid
development of new technologies to study the compositions of low-mass
extrasolar planets \citep[see, e.g., the review by][]{ks03} have
compelled several authors to consider planets with chemistries
unlike those found in the solar system \citep{stev04} such as water
planets \citep{kuch03, lege04}.  Here we describe a new possibility:
extrasolar planets in which carbon is the most abundant
component by number---carbon planets.

Recently, \citet{lodd04} argued that the {\it Galileo}-measured
abundances of CH${}_4$ and H${}_2$O in Jupiter's atmosphere imply that
O is depleted by a factor of 4 and C is enriched by 1.7 relative to
solar abundances, giving C/O = 1.8.  She suggested that the
planetary embryo that grew into Jupiter may have formed where the
nebula was locally carbon-rich, and that Jupiter's embryo was a carbon
planet.  This suggestion inspired our research; if Jupiter could have
formed from a carbon-rich embryo, we would expect that carbon-rich
embryos should be relatively common and occasionally observable as carbon
planets.  Although \citet{lodd04} did not provide the only possible
interpretation of the {\it Galileo} data, it seems reasonable that the
solar nebula and other planetary systems could have formed large
carbon-rich bodies.  This Letter discusses formation scenarios for
carbon planets (Section~\ref{sec:formation}), their likely compositions
(Section~\ref{sec:composition}), and their possible
appearances (Section~\ref{sec:appearance}).

\section{FORMATION OF CARBON PLANETS}
\label{sec:formation}

\subsection{Condensation}

If solar composition gas at $10^{-4}$ bars is cooled slowly from high
temperatures, several major building blocks of the solar system
condense out one by one.  First metal oxides and iron-peak elements
condense at $\sim$1500~K, then silicates condense at 1200--1400~K,
water at $\sim$180~K, and eventually, ammonia and methane at lower
temperatures \citep[e.g.,][]{lodd03}.  This equilibrium condensation
sequence apparently describes the gross compositions of the inner solar
system planets: Fe and Ni cores surrounded by silicate mantles, topped
by more complicated veneers containing water and more volatile
compounds. \citet{lewi74} connected these two trends, suggesting that
the locations of the planets determined their compositions; the solar
system terrestrial planets formed in hot regions of the solar nebula
from high-temperature condensates while the planets with larger
semi-major axes formed in cooler regions of the solar nebula from
lower-temperature condensates.  Some lower temperature condensates
spread around the planetary system via small bodies late in the
process of planet formation, coating planetary surfaces with
volatiles.  Although the details of this picture have
evolved---chemical processing is thought to be quenched in the outer
solar system, planets are now known to migrate, etc.---the equilibrium
condensation sequence survives as a standard reference point for
understanding the compositions of the solar system planets.

In gas with C/O ratio $>$ 0.98, the condensation sequence changes
dramatically \citep{lari75}.  In carbon-rich gas, the highest
temperature condensates ($T \approx 1200$--$1600$~K) are carbon-rich
compounds: graphite, carbides, nitrides, and sulfides.  Since the sun
has C/O ratio 0.5 \citep{aspl05}, carbon-rich condensation sequences
are not ordinarily associated with planet formation, though they have
been investigated at length in the context of the formation of silicon
carbide (SiC) grains in meteorites \citep[e.g.,][]{lodd95} and also in
the context of outflows from evolved stars \citep[e.g.,][]{lodd97,
lodd99}. 

Low-mass planets formed via these carbon-rich condensation
sequences would be carbon planets, initially composed largely of the high-temperature
condensates formed in carbon-rich gas, like graphite and silicon carbide.

\subsection{Formation Scenarios}

Some protoplanetary disks may spawn many carbon planets simply because they are
especially rich in carbon overall, and planet formation proceeds by a carbon-rich
condensation sequence.  The planets around the pulsar PSR 1257+12 \citep{wols92}
might have been formed in a carbon-rich nebula created by the disruption of either a
carbon-rich star or of a white dwarf \citep{tutu91, pods91, phin93}; perhaps
these pulsar planets are carbon planets.  These mechanisms may also operate
around white dwarfs \citep{livi92}.  In general, C/O ratios in stars and H II regions increase
with metallicity and towards the galactic center \citep[e.g.,][]{este05}, as reflected in 
galactic chemical evolution models \citep[e.g.,][]{gavi05}; planets detected by
microlensing may stand a better-than-average chance of being carbon
planets because they are closer to the galactic center than the Sun.
Stars that host extrasolar planets are on average enhanced in metals,
including carbon, and often show 10--15\% enhancements in C/O ratio
compared to the sun \citep{gonz01}.

The formation of carbon planets in our own solar nebula is also 
consistent with our current understanding of planet formation.
\citet{lodd04} suggested that the orbital radius in the solar nebula
corresponding to a temperature of 350~K should be called the ``tar
line'' by analogy to the snow line (the orbital radius where water ice
sublimates), and that a high concentration of carbon should accumulate
there as carbon diffuses outward from interior to that radius and as
carbon-rich grains spiral inwards toward that radius.  Condensation
lines like the tar line described by \citet{lodd04} now seem less
important for accumulating material than they used to, since new
calculations of the Prandtl numbers in turbulent disks imply that
outward chemical diffusion is likely too slow to play a major role
\citep{carb05}.  However, the inward spiral of solid grains probably
can substantially influence disk carbon chemistry.  \citet{youd04}
showed that over $10^5$ years, gas drag can augment the concentration
of solid grains in the central 2~AU of a protoplanetary disk by
factors of 10--100; this concentration of solids can trigger
gravitational instability of the solid layer in a quiescent
disk \citep{gold74}.  In such high concentrations, the material
imported by the grains
dominates the total C/O ratio in the disk.  Observations of
PAH features in Herbig Ae stars provide strong evidence for the
presence of carbon-rich grains in these disks.  If the grains in protoplanetary
disks have C/O ratios even slightly greater than 1, as some models of the spectra
of Herbig Ae stars suggest \citep{haba04}, the
particle-pileup mechanism of \citet{youd04} naturally leads to the
formation of carbon planets where the particles pile up.

A mechanism that could prevent this particle pileup process from
forming carbon-planets would be one that spreads condensates outward
from the hot center of the disk near the star faster than gas drag can bring in the
carbon-rich particles.  For example, some kinds of turbulence can
conceivably spread silicate grains around in a protoplanetary disk
faster than the carbon-rich grains pile up, as could X-winds \citep{shu01}.
However, the time-scale for gas
drag is reasonably short, and turbulent mixing is not likely to be
efficient throughout the disk, and the solid mass processed by winds
is uncertain and likely highly variable from system to system.  It is conceivable
that these processes that spread nebular solids were
effective at spreading solar composition equilibrium condensates
throughout the inner $\sim$2~AU of the solar system, but that outside
this region relatively carbon-rich bodies formed.

\subsection{Carbonaceous Chondrites and Fischer-Trope-Type Reactions}

Besides possibly Jupiter, carbonaceous chondrites demonstrate the presence of
planet-forming processes operating far from the chemical equilibrium
condensation sequence for solar-composition gas \citep[see the review
by][]{krot00}.  Carbonaceous chondrites are found with up to $\sim
6$\% carbon \citep{grad02} in the form of long-chain carbonaceous
compounds and smaller amounts of silicon carbide, graphite, and even
diamond.  This mass fraction of carbon greatly surpasses that of the
Earth, which is only $10^{-5}$--$10^{-4}$ carbon by mass
\citep{lodd98}.  It is likely that the more carbon-rich meteorites are
oxidized on impact and never recovered.  In standard equilibrium
condensation sequence for solar-composition gas, carbon in the solar
nebula remains entirely in the gas phase as CO at high temperatures
and CH$_4$ at low temperatures except where temperature drops below
$\sim 78$~K \citep{lewi72}.  At $\sim 78$~K, CH$_4$ becomes trapped in
water ice as a clathrate hydrate.  In this picture, we should expect
carbon to be found only as methane, mostly on the outermost planets or
in small quantities on the surfaces of inner planets, and always in
combination with large quantities of water from the ice that trapped
the methane.  Yet the parent bodies of the carbonaceous chondrites are
likely C-type asteroids that formed at much higher nebular
temperatures than 78~K.

\citet{krot00} reviewed some of the sources of solid carbon in
carbonaceous chondrite formation, describing three potentially
important processes:

\begin{trivlist}

\item{{\it Local C/O Enhancement}: Some regions of the solar nebula
may have had C/O ratios $> 0.98$, yielding the high-temperature carbon-rich
condensates described above. }

\item{{\it Direct Incorporation of ISM Carbon}: The presence of
pre-solar materials, including diamonds \citep[e.g.,][]{huss95}, in
meteorites demonstrates that some of the carbonaceous material from
the molecular cloud out of which the protoplanetary nebula formed was
preserved in the solar nebula and directly incorporated into
protoplanetary material.  Up to half of all carbon in the interstellar
medium exists as organic solids \citep{ehre00}, including amorphous
carbon, coal, soot, quenched-carbonaceous condensates, diamonds and
other compounds.  Larger carbon-bearing molecules such as polycyclic
aromatic hydrocarbons (PAHs), fullerenes, and long-chain carbon
compounds are present in ISM gas and incorporated into ISM dust grains
\citep{ehre00}.}

\item{{\it Fischer-Trope-type reactions (FTTs)}: Nebular grains may
undergo surface-mediated Fischer-Trope-type reactions, in which
transition metals promote the conversion of nebular CO and H$_2$ into
hydrocarbons at relatively high temperatures. }

\end{trivlist}

Besides the accumulation of imported carbon-rich ISM grains, FTTs may
also contribute to the formation of carbon planets.  Laboratory
evidence suggests that grain surfaces poisoned with FTT reaction
products also mediate FTTs; this cyclic process can conceivably
convert as much as half of the disk's gaseous carbon to solid organics
(J. Nuth 2004, private communication) at temperatures of $\sim$500~K
where FTTs are most efficient \citep{llor98,kres01}.

\section{COMPOSITION AND SURVIVAL OF CARBON PLANETS}
\label{sec:composition}

Our picture of the composition of carbon planets stems from the
equilibrium condensation sequence for C/O enriched nebular gas; this
analogy could be expected to extend to a wide variety of planetary
structures and compositions.  But now let us be more specific, and
hereafter refer to a planet as a carbon planet only if carbon is its
most abundant component by number.  This definition excludes
Jupiter-mass planets with carbon-rich cores \citep{lodd04} and
marginally excludes the extremely CO-rich gas giants that might be
found around pulsars or massive white dwarfs \citep{livi92}.  These
CO-rich gas giants could form in accretion disks created from
the disruption of a C-O white dwarf, disks that contain hundreds of
Jupiter masses of carbon and oxygen and scarcely any hydrogen.  They
would probably contain roughly equal parts carbon and oxygen.

Carbon planets, by our definition, are likely to be less massive
planets without massive gas envelopes to dilute the carbon content.
Such low-mass planets could form like terrestrial planets, as
carbon-rich embryos that were too small to acquire a gaseous envelope
and evolve to a gas giant planet.  The masses of such failed embryos
can range up to $\sim 60$~M${}_{\bigoplus}$ \citep{rafi04}.
Alternatively, carbon planets could form as gas-giant planets that
migrate inward and lose their gaseous envelopes. Gas envelope loss
close to the parent star could occur either by hydrodynamic escape
associated with stellar UV and extremely high exospheric temperatures
\citep[e.g.,][]{leca04} or by Roche-lobe overflow, a mechanism that
may help park migrating planets at small orbital periods
\citep{tril98,gu03, mcar04}.

Now we can return to the carbon-rich equilibrium condensation
picture.  This picture probably describes the gross internal
structure of carbon planets.  FTTs and direct incorporation of ISM
carbon can also help provide solid carbon enrichment, but the high
temperatures of planet accretion should process planetary interiors to
near chemical equilibrium after these non-equilibrium mechanisms would
operate.  In this equilibrium picture, carbon planets should be composed of
SiC, graphite and a lesser amount of other minerals that form in
a reducing environment \citep[see][]{fegl87, came88}, with cores made of iron
and iron-peak elements and surface layers of graphite and delivered
volatiles.   

Carbon condensation is strikingly different from the oxygen-rich
condensation sequence because of the availability of graphite as a
high-temperature condensate; there is no analogous high temperature
condensate of pure oxygen.  In a typical condensation sequence for
carbon enriched gas \citep{lodd97}, CO forms, using up all of the
oxygen, and then carbon left over from CO formation (called
``condensable carbon") condenses as SiC or TiC using up about half of
the Si \citep[e.g.,][]{lodd97}.  Most of the carbon left over after
SiC and TiC formation condenses as graphite; the higher the C/O ratio,
the more graphite forms.  Pure carbon can dissolve in metals and in
the planet's carbide mantle, but since graphite is less dense than SiC
(2 g~cm${}^{-1}$ compared to 3.2 g~cm${}^{-1}$), we might expect a
pure carbon layer to form on top of the SiC in a completely
differentiated planet.

Using this basic picture of planetary composition dictated by
carbon-rich condensation, \citet{fegl87} and \citet{came88} discussed
carbon planets in terms of what Mercury would have been like if it
formed in a carbon-rich environment. In their
discussion, graphite would be the predominant form of carbon by mass,
yielding a planet of lower density than silicate or iron planets.
They discuss the changes in planetary composition as a function of C/O
ratio, including the specific presence and absence of minerals found
in enstatite chondrites, the formation of Fe$_3$C and FeSi alloys
instead of pure Fe, and the presence of SiC and other unusual minerals
like CaS, TiN and AlN.

A possible concern about the existence of carbon planets is that they
could be threatened by collisional devolatilization.  However,
although carbon is usually considered a volatile in the solar system
this concern about carbon planets is likely unwarranted.  On carbon
planets, large amounts of carbon can be locked into refractory
compounds, as oxygen is on Earth.  For example, SiC is a sturdy
ceramic used for lining the cylinders of motorcycle engines; it
remains solid until temperatures of $\sim 2800$~K at pressures of 35
bars \citep{tair88}.  Although graphite should emerge as a surface
layer in a differentiated carbon planet, a few km into the planet's
interior at a pressure of $\sim 10^5$ bars, pure carbon in a cool
carbon planet should turn to diamond.  Diamond remains solid up to
temperatures of at least $4000$~K under a thin atmosphere
\citep{grum96}.  If the pure carbon in a carbon planet does not
completely partition into the core or mantle, the resulting diamond
shell could protect the planet from strong stellar radiation even as
close $0.03$~AU from a solar-type star, where other volatiles would
easily be stripped.

\section{APPEARANCE OF EXTRASOLAR CARBON PLANETS}
\label{sec:appearance}

Since all we can see of a planet is its surface and
atmosphere, it is difficult to confidently ascertain the bulk
composition of a planet---this holds true even for planets in
our own solar system.  However, under some circumstances, carbon
planets may contain distinguishing spectral features that we can
recognize from far away.  The low densities expected for carbon planets,
between those of water planets \citep{kuch03, lege04} and silicate planets,
may provide one more clue to their presence.

We will discuss the potentially recognizable properties of carbon planets in two rough
size ranges that we will refer to as ``Earth mass'' and ``Neptune
mass''.  By Earth-mass planets ($\lesssim 10$~M${}_{\bigoplus}$) we
mean those with thin secondary atmospheres like the terrestrial
planets in the solar system.  These planets have too low a mass to
retain the primary hydrogen atmospheres they acquire hydrostatically
from their protoplanetary disks.  By Neptune-mass
planets, we mean $\sim$10--60 $M_{\bigoplus}$ planets that retained thick
primary atmospheres like those of Uranus and Neptune, planets
whose rocky surfaces are too deep to see at any wavelength.

\subsection{Neptune-Mass Carbon Planets}

Neptune-mass carbon planets can have atmospheres massive enough not to
be dominated by volatiles delivered by late impacts of comets and
asteroids; their atmospheres are probably mostly H, like the
atmospheres of Uranus and Neptune.  However, since the atmospheres of
these Neptune-mass carbon planets would have formed in contact
with a large reservoir of carbon they likely have large C/O ratios.
Pollution by late carbon-rich impactors could conceivably generate
a high atmospheric C/O ratio on a planet that is not a carbon planet,
so finding a planet with a carbon-rich atmosphere does not guarantee
that the planet is a carbon planet.  But we can begin to understand
the appearance of a Neptune-mass carbon planet by investigating a
Neptune-mass planet with a large atmospheric C/O ratio.

We explored the equilibrium chemistry of a Neptune-mass planet's
atmosphere using a Gibbs free energy minimization chemical equilibrium
calculation \citep{seag99, seag00}.  We modeled atmospheres with the H
and He fractional abundances like those of Neptune's atmosphere (H$_2$
= 80\%, He = 19\%).  We considered two cases: an atmosphere with
metals in the same relative abundances as found in the solar
photosphere, (e.g., C/O = 0.5) and an atmosphere similar to the first
but with C/O = 1.01.

\begin{figure}
\plotone{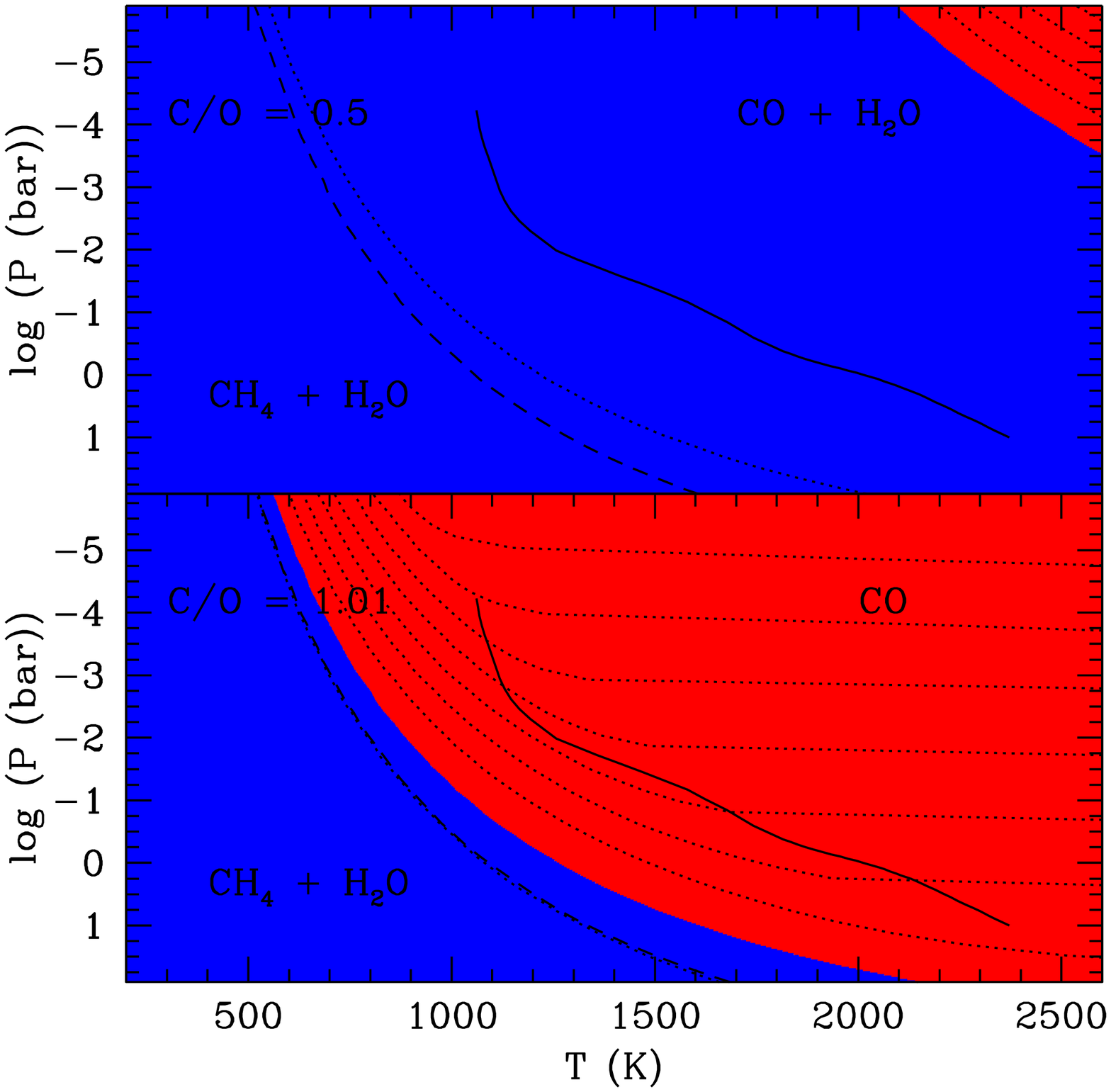}
\caption{Summary of the CO, CH$_4$, and H$_2$O concentrations at
chemical equilibrium in a Neptune-mass planet atmosphere with C/O $\approx
0.5$ (upper panel) and C/O $\approx 1.01$ (lower panel).  The blue
regions shows where H$_2$O/CO $>$ 0.1; red regions are drier.  Dotted
lines, from left to right, show where H$_2$O/CO=$1, 0.1, \cdots,
10^{-8}$.  Heavy dashed curves show where CO and CH$_4$ gases have
equal abundance; CO is the dominant carbon-bearing molecule at high
temperatures and CH$_4$ is the dominant carbon-bearing molecule at low
temperatures.  In the lower panel the dashed curve and leftmost dotted
curve overlap.  The solid black line shows a possible
temperature-pressure profile of a $T_{eff}=1500$~K hot Neptune
planet.}
\label{fig:chem}
\end{figure}

Figure~\ref{fig:chem} summarizes the equilibrium chemistry for CO,
CH$_4$, and H$_2$O as a function of temperature and pressure for these
two cases.  In the blue regions, H$_2$O is abundant, at $\geq 0.1
\times $ the CO number density.  In the red regions, H$_2$O is
relatively scarce.  A series of dotted lines in each panel indicate
contours of the H${}_{2}$O concentration at ${\rm H}_{2}{\rm O}/{\rm CO}=1,
0.1, \cdots, 10^{-8}$.  At high temperatures, CO is
the dominant carbon-bearing molecule, and at low temperatures CH$_4$ is
the dominant carbon-bearing molecule; thick dashed lines indicate the
temperatures and pressures where CO and CH$_4$ have equal abundances.

Cold planets will have temperature-pressure profiles residing largely
in the left side of Figure~\ref{fig:chem}, where H$_2$, CH$_4$ and
H$_2$O dominate in both the C/O=0.5 case and the C/O=1.01 case.  So
cold Neptune-mass planets will have similar dominant spectral
absorption features whether they have C/O = 0.5 or C/O = 1.01.
Planets with C/O ratios even greater than 1.01 will have a similar
chemical equilibrium profile to those with C/O = 1.01, though planet
atmospheres richer in carbon will have lower H$_2$O/CO ratios, higher
CH$_4$/H$_2$O ratios and a CH$_4$-dominated region at slightly
higher temperatures. The cold Neptune-mass carbon planets, therefore,
will be difficult to recognize as carbon planets because their spectra
should be very similar to other planets with solar-system abundances;
though they may be carbon enriched, their spectra should be dominated
by CH$_4$ and H$_2$O.  The limited spectroscopic capabilities
expected for extrasolar planet observations may make identification of such
cold carbon planets impossible.

Hot Neptune-mass carbon planets, on the other hand, will have
temperature-pressure profiles stretching into the right side of
Figure~\ref{fig:chem}, where, as Figure~\ref{fig:chem} shows, the
atmospheric chemistry is a strong function of C/O ratio.  These
planets will have a low atmospheric abundance of H$_2$O, the molecule
that normally dominates the spectra of hot planets.  In atmospheres
of hot planets with C/O $< 1$,  H$_2$, H$_2$O and CO molecules
dominate the chemistry.
For hot planets with C/O $>1$, H$_2$ is still the most abundant
atmospheric molecule, but CO is the main O-bearing molecule and
H$_2$O is scarce.

\begin{figure}
\plotone{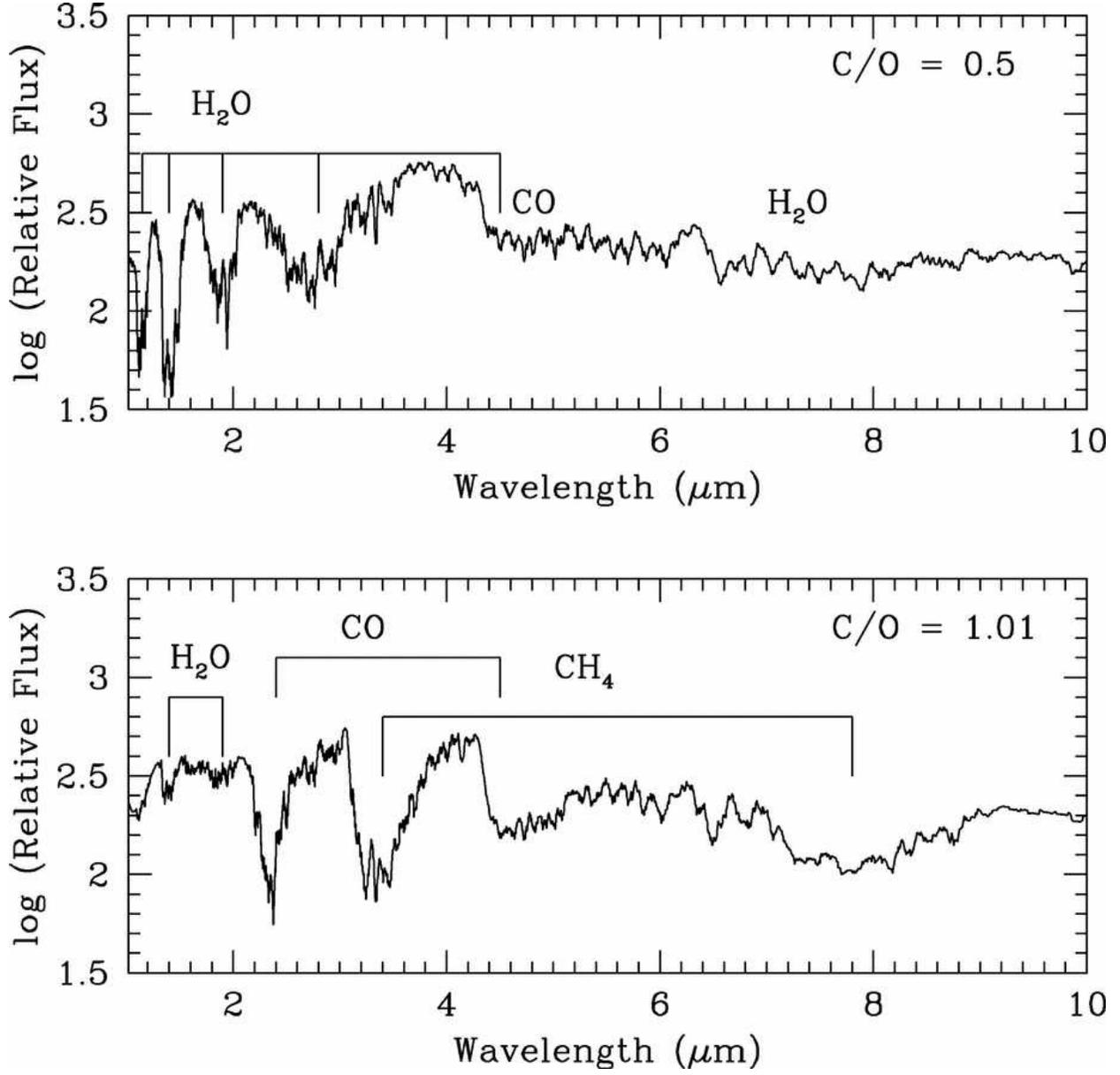}
\caption{Model thermal emission spectra for a hot Neptune-mass planet.
Upper panel: a C/O = 0.5 hot Neptune-mass planet atmosphere is
dominated by H$_2$O spectral features with some weak CO absorption
features. Lower panel: a C/O = 1.01 hot Neptune carbon planet
atmosphere has very little H$_2$O; instead CO and CH$_4$ absorption features dominate the spectrum.}
\label{fig:spectra} 
\end{figure}

Figure~\ref{fig:spectra} shows the significant differences between the
spectrum of a hot Neptune-mass planet with C/O = 0.5 and the spectrum of a hot
Neptune-mass carbon planet with C/O = 1.01.  We computed the spectra
with a 1D radiative transfer code described in \citet{seag05}.  We
computed the temperature-pressure profile using a 1D
radiative-transfer/radiative equilibrium/hydrostatic equilibrium code
\citep{seag00, seag05} for a cloud-free planet with Neptune's mass and
radius and C/O=0.5 located 0.04 AU from a G8V star.  This
temperature-pressure profile appears in Figure~\ref{fig:chem} as a
solid curve in each plot.  Although atmospheres of different
compositions should have somewhat different temperature-pressure
profiles because of their different interactions with stellar
radiation, for this exploratory calculation we made the approximation
of using the same profile for both synthetic spectra.

Besides water abundance, a second likely major difference between hot
carbon planets and hot solar-system-abundance planets is the grain
chemistry.  The solar-system abundance planets are likely to have Fe
clouds and silicate clouds, whereas the carbon planets are likely to
have Fe clouds, SiC clouds, and graphite clouds.  Graphite clouds can
be dark---much darker than silicate clouds.  Graphite clouds tend to
deplete carbon until the gas C/O ratio is near 1 \citep{lodd97}; this
possibility is why we assumed a
C/O ratio only slightly greater than 1 in our carbon-rich models.  In
a carbon-rich atmosphere containing hydrogen and some iron grains or
other suitable catalyst, atmospheric FTTs can potentially slowly
convert CO into methane.  These details will need to be incorporated
into a more sophisticated carbon-rich model atmosphere.

The reducing atmospheres of cold Neptune-mass carbon planets should
also allow for photochemical hydrocarbon synthesis, as the atmospheres
of solar system giant planets do \citep[e.g.,][]{stro83,glad96}.  For
example, the atmospheres of all the giant planets in the solar system
contain ethane (C${}_{2}$H${}_{6}$) and acetylene (C${}_{2}$H${}_{2}$)
identifiable via infrared bands at 3.3--3.4~$\mu$m and
11.6--12.4 $\mu$m (ethane) and 2.9--3.2~$\mu$m and 13.7~$\mu$m
(acetylene) \citep{roth92}.

\subsection{Earth-Mass Carbon Planets}

Carbon planets with little or no atmosphere may be even harder to
recognize spectroscopically than the cold Neptune-mass carbon planets.
Such low-mass carbon planets could gave undergone dramatic
atmospheric evolution and therefore will have a wide variety of
appearances, comparable to the variety of silicate planets seen in the
solar system (i.e., Earth, Venus and Mars).  

Earth-mass carbon planets would
probably not have atmospheric CO$_2$, the dominant atmospheric
molecule on Mars and Venus.  The surface layers of terrestrial planets,
including their atmospheres, are shaped by the delivery of low-temperature
condensates that were not processed to chemical equilibrium.  A variety of
delivery agents may be important \citep[e.g.,][]{kyb04}, but we can take a
comet as an example.  A comet contains both graphite and silicates and metal
oxides in its core, materials that would combust if heated, releasing CO.
When a comet arrives at a terrestrial planet, it meets an oxygen-rich
environment where water and silicates can survive while the delivered
carbon is easily oxidized.  When a comet arrives at a carbon planet, it
meets a reducing environment, and a carbon-rich
surface to stir up and chemically interact with.  Here delivered carbon can
survive, while the water is easily converted to CO, H, and hydrocarbons, and the
silicates are easily converted to carbides.  So the hallmark of a
terrestrial-mass carbon planet's atmosphere might be the absence of
oxygen-rich gases like CO${}_{2}$, O${}_2$, O${}_3$, etc., and the
dominance of CO, or, on a cold carbon planet, CH${}_{4}$.

Some Earth-mass carbon planets in short-period orbits may have lost most or all of their atmospheres from atmospheric escape, revealing their solid surfaces.  Such planets would be stable near their host stars, protected by graphite, diamond, or silicon carbide
shells.  We would expect carbon planets with no atmosphere to be dark, with
relatively featureless spectra at visible wavelengths like the
graphite-rich rock of carbonaceous chondrites, since an admixture of a
few percent of small carbon grains in a rock that would otherwise be
light-colored readily damps spectral features from other minerals
\citep{clar83}.  

Cold low-mass carbon planets could also foster
the long-term survival of photochemically-synthesized long-chain carbonaceous
compounds that would be burned to CO in more oxidizing atmospheres.  These compounds can
rain onto the planet's surface, creating tar oceans like those
proposed for Titan \citep{hunt92,luni93}.   So a carbon planet that
has lost its atmosphere could conceivably retain a coating of
long-chain carbon compounds, i.e., tar.

\section{SUMMARY AND DISCUSSION}
\label{sec:conclusion}

A logical companion to the picture championed by \citet{lewi72} and
others for planet formation via the equilibrium condensation sequence
for solar composition gas is planet formation via equilibrium
condensation of carbon-rich gas.  A mere factor of 2 increase in the
C/O ratio beyond solar leads to a different new suite of carbon-rich
high temperature condensates.  We briefly explored some
planet-formation scenarios that would lead down this path and
discussed their likely products, carbon planets composed largely of
SiC, graphite, and lesser amounts of other carbides and minerals
that form in a reduced environment.  These planets would also likely
contain metal cores and possibly contain graphite or diamond layers.

We also discussed the observational signatures of carbon planets.  
The CO-dominated spectra of hot Neptune-mass carbon planets should 
distinguish them from the H${}_2$O-dominated spectra of hot Neptune-mass
planets with solar-composition atmospheres (Figure 2).  Cold carbon planets
of all sizes could have stable long-chain hydrocarbons in their atmospheres
and on their surfaces that would be destroyed in a more oxidizing environment.
Diamond or graphite shells can protect carbon planets
from devolatilization by heat and ionizing radiation as close as a few
stellar radii from a solar-type star.

Planets around pulsars and white dwarfs are good candidates for carbon
planets; they could have formed in a carbon-rich nebulae created by the
disruptions of carbon-rich stars or white dwarfs.
Carbon planets should be more common around more metal-rich stars and
stars near the Galactic center such as microlensing targets.  Given the slow
increase in Galactic C/O ratio with time, there may even
come a day when the C/O ratio of nearby star forming regions is so high that
all new planets formed in the solar neighborhood are carbon planets.  Our
current picture of planet formation is also consistent with carbon
planets forming in protoplanetary disks like the solar nebula;
particle pileups \citep{youd04} of carbon-rich grains from the ISM can
create the needed carbon-rich environment.

Carbon planets may soon regularly turn up in transit studies and in searches
for extrasolar Earth analogs.  The possibility
of low-mass carbon-planets suggests that a key element of
characterizing planets with the Terrestrial Planet Finder (TPF)
missions should be determining whether their atmospheres are oxidizing
or reducing.  It is tempting to rely on the presence of CO${}_{2}$ as
a hallmark of terrestrial planets; but if low-mass carbon planets are
common, many small planets may not share this feature.

What other possible kinds of planets are there?  The planet zoo now
contains silicate planets (e.g., Earth and Mars), hydrogen and
helium planets (e.g., Jupiter and Saturn), water planets (which
perhaps we might think of as oxygen planets), iron planets
\citep{stev04}, and carbon planets.  A glance at a table of solar
abundances suggests that next we might consider helium, neon, and
nitrogen planets.  Neon seem disadvantaged as planet-building material
because of its chemical inertness.  However, some AGB stars could have
envelopes greatly enriched in nitrogen \citep{domi84}, possibly
suggesting a mechanism for forming nitrogen-enriched planets.

Could there be life on a carbon planet?  Life on Earth appears to owe
its existence to the presence of carbon and water and other volatile
compounds delivered late in the planet-formation process in trace
quantities.  It is reasonable to think that the chemical microcosm
required to support life could likewise be delivered to the surface of
a terrestrial-mass carbon planet.  One thing seems likely---if a
carbon planet hosted intelligent creatures, they would not go to war
over such a common trifle as carbonaceous fuels.

\acknowledgements

Thanks to Katharina Lodders, Bruce Fegley, Joe Nuth, Roland Heck, and
John Chambers for helpful conversations.  M.J.K. acknowledges the
support of the Hubble Fellowship Program of the Space Telescope
Science Institute. S.S.  is supported in part by NAG5-13478, by NASA
NAI-CAN NA04CC09A, and by the Carnegie Institution of Washington.

\end{document}